# Micrometer-scale monolayer SnS growth by physical vapor deposition†

H. Kawamoto[a], N. Higashitarumizu[a], N. Nagamura[b,c], M. Nakamura[b], K. Shimamura[b], N. Ohashi[b], and K. Nagashio[*a]

[a]Department of Materials Engineering, The University of Tokyo, Tokyo 113-8656, Japan
[b]National Institute for Materials Science, Ibaraki 305-0044, Japan
[c]PRESTO, Japan Science and Technology Agency, Saitama, 332-0012, Japan

[*]nagashio@material.t.u-tokyo.ac.jp

†Electronic supplementary information (ESI) available

**Abstract:**
Recently, monolayer SnS, a two-dimensional group IV monochalcogenide, was grown on a mica substrate at the micrometer-size scale by the simple physical vapor deposition (PVD), resulting in the successful demonstration of its in-plane room temperature ferroelectricity. However, the reason behind the monolayer growth remains unclear because it had been considered that the SnS growth inevitably results in a multilayer thickness due to the strong interlayer interaction arising from lone pair electrons. Here, we investigate the PVD growth of monolayer SnS from two different feed powders, highly purified SnS and commercial phase-impure SnS. Contrary to expectations, it is suggested that the mica substrate surface is modified by sulfur evaporated from the $Sn_2S_3$ contaminant in the as-purchased powder and the lateral growth of monolayer SnS is facilitated due to the enhanced surface diffusion of SnS precursor molecules, unlike the growth from the highly purified powder. This insight provides a guide to identify further controllable growth conditions.

## 1. Introduction

Tin monosulfide (SnS), which is a group IV monochalcogenide (SnX, GeX; X: S, Se), has been widely studied for low-cost and scalable thin-film photovoltaics due to a theoretical power conversion efficiency close to the Shockley-Quiesser limit of ~32%[1-3] and for thermoelectric applications due to its large figure of merit ($ZT$ = ~0.6).[4,5] SnS is also a member of the relatively unexplored layered two-dimensional (2D) semiconductor family and is isostructural to black phosphorus.[6-8] Recent theoretical predictions of giant piezoelectricity and ferroelectricity[9-13] have facilitated their experimental investigation. These properties rely on the breaking of inversion symmetry in odd-numbered layers, which results from the crystal structure consisting of two elements with different electronegativities, unlike black phosphorus. To prove the piezoelectric and ferroelectric properties, a monolayer is highly required since these properties become prominent in few-to-monolayer SnS with reducing thickness. In contrast to other typical 2D materials,[14] mechanical exfoliation of monolayer SnS has not been achieved[15-17] because the interlayer interaction is strong due to the lone pair electrons in the Sn atoms, which generate a large electron distribution and electronic coupling between adjacent layers.[18-20]

The synthesis of high-quality monolayer SnS with a large area suitable for device fabrication is crucially important to prove the predicted properties. **Figure 1** summarizes the typical growth modes observed for SnS growth thus far.[1,2,21-25] On reactive metallic substrates typically used in photovoltaic research (a), SnS layers are inclined at large angles to the substrate surface due to the chemisorbed edge bonding of SnS.[1,2,22] On the other hand, the basal-plane growth by physisorption (b) was realized on nonreactive substrates because a larger area of SnS contacting to the substrate was energetically favorable.[21,23-25] However, in-situ observation[21] during basal-plane growth clearly elucidated that vertical growth of SnS is facilitated because the surface diffusion of precursor molecules is limited on nonreactive substrates and the adsorption on the SnS basal plane is relatively strong due to the lone pair electrons, which explains why the monolayer is limited to a few hundred nanometers in diameter[21,26] and only thicker multilayer SnS is accessible for characterization of physical properties, such as optical properties,[24,27] the anisotropic crystal structure associated Raman spectrum[28,29] and electrical transport,[30] valley polarization,[31,32] and ferroelectricity.[26] Nevertheless, micrometer-scale monolayer SnS was recently realized on a mica substrate by simple physical vapor deposition (PVD) and its in-plane room temperature ferroelectricity was successfully demonstrated by the switching behavior in the round sweep *I-V* characteristics.[33] The reason behind the monolayer growth remains unclear.

Looking back on past photovoltaic research,[3] purification of commercial feed powder has been reported to drastically improve the power conversion efficiency owing to the exclusion of the phase impurities $Sn_2S_3$ and $SnS_2$. Therefore, two kinds of feed powders are studied in the present work: commercial SnS powder (3Nup, Kojundo Chemical Laboratory Co.) and a SnS powder crushed from high purity single crystal SnS, which was grown by the horizontal gradient freeze method from S powder purified by distillation and Sn shot (4N).[34] These are hereafter referred to as commercial powder and high purity powder, respectively. Their powder X-ray diffraction (XRD) patterns in **Fig. 2** clearly show the phase impurities $Sn_2S_3$ and $SnO_2$ in the commercial powder but not in the high purity powder.

In this article, we address monolayer SnS growth on an atomically flat mica substrate from different powders to identify the mechanism for micrometer-scale monolayer growth. Since chemical characterization of the nanoscale microstructure is determinant in this study, a core-level photoelectron spectromicroscopic apparatus installed at the synchrotron radiation facility of SPring-8, called "3D nano-ESCA,"[35,36] is utilized, with which we can scan the sample with a high lateral spatial resolution of ~70 nm to record photoelectron spectra for quantitative analysis of the chemical states.

## 2. Results & discussion

The SnS was grown on a mica substrate with $10\times10\times0.5$ mm[37] in a three-zone heating tube furnace by PVD, which is schematically shown in **Fig. S1a**. As a preliminary experiment, the amount of SnS powder consumption was monitored during the growth run in a $N_2$ atmosphere. It did not linearly decrease with the growth time and was saturated roughly at 10 min. Therefore, the growth substrate was first heated to the predetermined temperature, then, the feed SnS powder was heated, and the growth time was limited to less than 10 min, as shown in **Fig. S1b**. **Figure 2** compares the XRD data for the high purity and commercial powders before and after the growth run. For the high purity powder, only SnS is detected both before and after the growth run. For the commercial powder, on the other hand, $Sn_2S_3$ disappeared after the growth run while $SnO_2$ remained. This can be understood from the fact that the equilibrium vapor pressures of $SnS(g)$ and $S(g)$ are much higher than those of $SnO(g)$ and $SnO_2(g)$ based on a simple thermodynamic calculation of the equilibrium vapor pressures for the reactions,[38,39] as shown in **Fig. S2**.

To suppress the vertical growth rate of SnS, the three-zone furnace temperatures and growth pressure were systematically investigated.[33] Finally, thin SnS films were grown with various thicknesses on mica substrates at $T_R = 410°C$, $T_C = 440°C$, $T_L = 470°C$ and a growth pressure of 10 Pa. **Figures 2b and c** show atomic force microscopy (AFM) topographic images of typical SnS films grown from the high purity powder and commercial powder, respectively. Both are relatively thin SnS films (~2.8 nm for the high purity powder and ~1.2 nm for the commercial powder) compared with the monolayer thickness of ~0.6 nm. To compare the crystallinity, the Raman spectra for the SnS films with the similar thickness (~4 nm) are shown in **Fig. 2d**. For both crystals, only the specific Raman peaks of SnS were confirmed,[28,29,33] excluding a possibility of other phases: $SnS_2$ and $Sn_2S_3$. Contrary to expectations, the Raman intensities of the SnS film grown from the commercial powder is higher than that from the high purity powder, suggesting that the crystallinity for the commercial powder is superior to that for the high purity powder. The SnS film grown from the high purity powder shows a rounded rectangular shape reflected from the diamond shape with thermodynamically stable facets for bulk SnS, while the shape of the SnS film grown from the commercial powder is random circular. According to the classical growth theory,[40] when the growth proceeds with time, the planes with faster growth rate initially grow preferentially. Finally, the crystal is surrounded by the thermodynamically stable facet planes whose growth rate is lowest. However, when thickness is quite thin, the thermodynamically stable shape has not been reached. This is case for the rounded shape of SnS. Moreover, only for the commercial powder, the random circular shape was found, indicating that the thickness reached from commercial powder is thinner than that from the high purity powder. On the other hand, atomically flat submicron grains with heights similar to those of monolayer SnS are observed around the SnS film only for the commercial powder, but not for the high purity powder. These grains could be different from SnS because the SnS Raman peaks are not detected in region (ii) and the spectrum is consistent with that of a bare mica substrate with a peeled surface. Despite the phase-pure high purity feed powder, SnS films thinner than ~2 nm were not observed even though different growth conditions were explored.

Interestingly, the micrometer-scale monolayer SnS is only found when using the commercial powder, as explained below. This suggests that the atomically flat submicron grains could be key for the enhanced lateral growth.

Therefore, SnS growth from the commercial powder is addressed here. **Figure 3a** shows an AFM image of another SnS film on a mica substrate, where the submicron grains can be seen below as well as around the SnS film, schematically shown as type II in **Fig. 3c**. This is confirmed by the phase image (ii), in which the SnS film is identified as a continuous film. Although it is well known that the phase image contrast changes with the layer number even for the same substance,[41] the fact that the contrast of the SnS film on the submicron grains is the same as that on the mica substrate clearly suggests that the submicron grain is a different substance than SnS. Similar images can be found in **Fig. S3a**. The height of the submicron grain at (1) is ~0.51 nm, which is slightly lower than the height of ~0.85 nm for SnS film at (2) in **Fig. 3a**. Moreover, this sample was reheated at 400°C for 20 min in $N_2$ atmosphere without SnS feed powder. The SnS film desorbed, while the submicron grains mostly remained, as shown in (iii) of **Fig. 3a**. The detailed method is explained in **Fig. S3b**. These results also support that the submicron grain is a different substance than SnS. Here, the submicron grains often uniformly cover the entire mica substrate, and the grain boundaries cannot be identified in the AFM image, which leads to the miscounting the layer number of the SnS film. One example is shown in **Fig. 4**. Although the AFM image of **Fig. 4a** suggests a monolayer SnS film with a thickness of 0.8 nm, a high-angle annular dark-field scanning transmission electron microscopy (HAADF-STEM) image at position (i) in **Fig. 4b** proves that it is indeed a bilayer SnS film. That is, type III is the case for the SnS film in **Fig. 4a**. Based on this observation, the SnS films in **Fig. 2c** and **Fig. 3a** can be similarly identified as a bilayer on mica (type III) and a monolayer on submicron grains (type II), respectively. Note that type III is easier to appear than type III from the experimental observation. Monolayer SnS directly on mica (type I) is shown in **Fig. 3b**. In fact, this SnS film had been confirmed to be monolayer by cross-sectional TEM by adjusting the zone axis using the polarized Raman spectroscopy in the previous research.[33] Although the submicron grains are slightly observed, the thickness of SnS film measured from the surface of the mica substrate is indeed ~0.9 nm. The holes are often observed especially for monolayer, because the growth at the low pressure of 10 Pa enhanced the SnS desorption during the growth. Therefore, these are probably etch pits created during the growth. When a post-growth annealing was performed for the SnS crystals, aligned square-shaped etch pits were clearly observed, indicating a single crystalline nature. A more representative morphology of monolayer SnS surrounded by submicron grains can be seen in Fig. 1d in ref. 33.

To characterize the submicron grains, the HAADF-STEM image at position (ii) is shown in **Fig. 4c**. As expected, a continuous feature composed of the aggregation of submicron grains is detected on the mica, which is different from that for the SnS film. The composition ratio (at%) of Sn : S for continuous submicron grains in the rectangular region is determined to be 1:1.36 using energy dispersive spectroscopy (EDS). Although other additional signals from Ga, F, Mg, Si and Al are also detected, as shown in **Fig. S4**, these signals are attributed to the contamination from the mica substrate and focused ion beam processes during the TEM sample fabrication. Since the composition ratio for Sn:S in bilayer SnS is analyzed as 1:0.81 by EDS, the submicron grains could be $Sn_2S_3$, not $SnS_2$.

To prove that the submicron grains are indeed $Sn_2S_3$, scanning photoelectron microscopy measurement using 3D nano-ESCA was carried out. **Figure 5a** shows an optical image of the SnS/mica sample specially prepared for 3D nano-ESCA. To prevent charge-up on the insulating mica substrate during the ESCA measurement, patterned Ni/Au electrodes were fabricated via standard electron beam lithography and metal evaporation. The ground wire was extended to the 3D nano-ESCA sample holder with Ag paste and Cu wire, as shown in **Fig. S5**. **Figure 5b** shows an intensity mapping image for the Sn $3d_{5/2}$ peak with a spatial resolution of 100 nm. The pinpoint Sn $3d_{5/2}$ peak core-level spectra recorded at points of the (i) inner intrinsic SnS region and (ii) submicron grains around the SnS film are shown in **Fig. 5c**. Although analysis of the ESCA spectra on the three different phases, SnS, $Sn_2S_3$, and $SnS_2$ has been debated mainly due to the difficulty of preparing phase-pure samples, especially $Sn_2S_3$,[42,43] a recent study clearly identified the spectra by preparing each single crystal.[44] Here, the present spectra are analyzed by following the previous identification. Using multicurve fitting, a single main peak and a small subpeak were found for point (i). The peak position for the main peak is approximately 485.9 eV, which corresponds to Sn(II). The small subpeak could be assigned to the single oxidation state of Sn(II). At point (ii), the main peak of Sn(II) was also confirmed. In addition to the Sn(II) peak, a strong peak was found at approximately 486.5 eV, which corresponds to Sn(IV). Since $Sn_2S_3$ is the mixed-valency compound between SnS and $SnS_2$, this assignment is strengthened by the fact that the area ratio was calculated to be 1:1. These results indicate that the inner region of the large film at point (i) is SnS, while the outside region at point (ii) is indeed $Sn_2S_3$.

Finally, let us consider the role of submicron $Sn_2S_3$ grains in the micrometer-scale lateral growth of monolayer SnS film. These $Sn_2S_3$ grains are peculiar to the commercial powder, which initially contains a small amount of $Sn_2S_3$. Although monolayer SnS often grows on $Sn_2S_3$ grains (**Fig. 3a**), this is not always the case (**Fig. 3b**). S-rich conditions on the mica substrate at the early stage of the growth run, rather than $Sn_2S_3$ itself, possibly contribute to the monolayer growth. Recently, it was reported that an increase in the sulfur vapor concentration in the chemical vapor deposition of $MoS_2$ leads to alignment of $MoS_2$ on a *c*-plane sapphire substrate.[45] Moreover, STEM observations suggested that the *c*-plane sapphire surface is terminated by Se during the epitaxial growth of $WSe_2$.[46] Therefore, one of the possible explanations is as follows. The sulfur could evaporate from the $Sn_2S_3$ contaminant at temperature lower than the SnS powder and cover the mica substrate during the heating of the powder. And then, the surface diffusion of SnS precursor molecules is facilitated at the early stage of the growth, as schematically shown in **Fig. 1c**. Although the lateral size of the SnS film grown from the high purity powder (**Fig. 2b**) is larger than that from the commercial powder (**Fig. 2c**), the thickness in **Fig. 2b** is thicker than that in **Fig. 2c**. If the lateral sizes with the same thickness are compared, SnS grown from the purchased powder is larger than that from the high purity powder. Based on the above discussion, we considered to grow monolayer SnS by two types of controlled methods. One is to add the $Sn_2S_3$ powder in the high purity SnS powder to facilitate the surface diffusion of SnS precursor molecules. However, $Sn_2S_3$ is thermodynamically metastable and is not commercially available. Therefore, as an alternative method, the S powder was added in the high purity SnS powder. However, because it is quite difficult to precisely control the amount of sulfur vapor due to the low sublimation temperature, we did not succeed. Nevertheless, surface modification of mica substrates by sulfur vapor will be key to further enabling scalable growth for device applications. Indeed, qute recently, the enhancement of lateral growth size was reported by excess sulfur even though the thickness is not monolayer but mutilayer.[47]

## 3. Conclusions

The reason why micrometer-scale monolayer SnS was grown by simple PVD was explored by comparing the growth from high purity and commercial powders. Contrary to expectations, monolayer SnS was grown only from the commercial powder. The submicron grains around the SnS film on the mica substrate were identified as $Sn_2S_3$ by 3D nano-ESCA, which suggests that the mica substrate surface is modified by sulfur evaporated from the $Sn_2S_3$ contaminant in the as-purchased powder and that the lateral growth of monolayer SnS is facilitated due to the enhanced surface diffusion of SnS precursor molecules, unlike the growth from highly purified powder. This insight obtained in this study provides a basis for identifying further controllable growth conditions.

## 4. Experimental
**Growth**
SnS crystals were grown by a home-built PVD growth furnace with three heating zones, as shown in **Fig. S1**. A commercially available SnS powder (3Nup, Kojundo Chemical Laboratory Co.) and high purity SnS powder[34] were used as sources. Since the particle size of the commercial powder is 100 ~ 200 µm, the particle size of the high purity powder is also reduced by the mortar and paste to the comparable size of 100 ~ 400 µm to obtain the same growth condition. To promote lateral growth, we used a freshly cleaved mica substrate sized 1 cm × 1 cm × 0.5 mm, whose surface is atomically flat. $N_2$ carrier gas was introduced into the furnace through the mass flow controller and the growth pressure was reduced to 10 Pa by a vacuum system to enhance the SnS desorption during the growth.

**Measurements**
Raman spectra were measured using a 488 nm excitation laser, whose penetration depth is ~20 nm in SnS. To avoid degradation of SnS during the measurement, the samples were measured in the vacuum at 5 K. The nominal $1/e^2$ spot diameter and laser power just below the objective lens were 2.5 µm and 0.5 mW, respectively. HAADF-STEM images were taken at the acceleration voltage of 200 kV using JEM-ARM200F to confirm the layer number and crystallinity. For a core-level photoelectron spectromicroscopy, 3D nano-ESCA installed at the soft X-ray beamline BL07LSU in the synchrotron radiation facility of SPring-8 was used. The photon energy of the incident beam was 1000 eV. To reduce the charge up during the measurement, the Ni/Au electrodes were prepared through the standard electron beam lithography technique, as shown in **Fig. S5**.

**Acknowledgments**
The authors acknowledge Y. -R. Chang for his support. N. H. was supported by JSPS KAKENHI Grant-in-Aid for JSPS Fellows Number JP19J13579, Japan. This research was supported by Samco Science and Technology Foundation, Yazaki memorial foundation for science and technology, The Canon Foundation, the JSPS Core-to-Core Program, A. Advanced Research Networks, the JSPS A3 Foresight Program, JSPS KAKENHI Grant Numbers JP19H00755, 19K21956, 18H03864, and 19H02561, MEXT Element Strategy Initiative to Form Core Research Center, Grant Number JPMXP0112101001, and PRESTO (Grant number: JPMJPR17NB) commissioned by the Japan Science and Technology Agency (JST), Japan. The spectral datasets were obtained with the support of the University of Tokyo outstation beamline at SPring-8 (Proposal Numbers: 2018B7580 and 2019A7451).

**Figure**

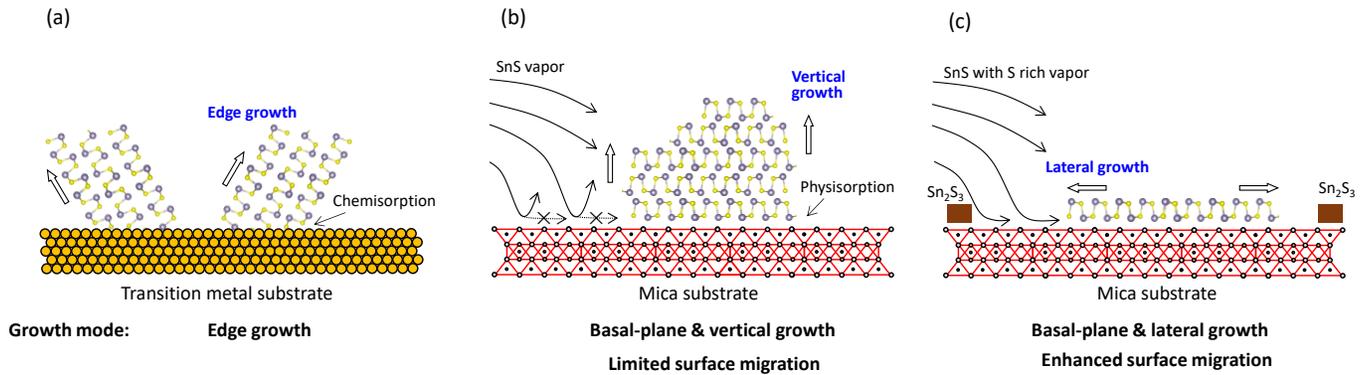

**Fig. 1** Typical growth modes observed for SnS growth. (a) Edge growth mode on reactive metallic substrates by chemisorption. Although covalent bonding between S and the metallic substrate is intuitively expected, it is not clear at present. Therefore, the elements bonding to the substrate are not selectively drawn. (b) Basal-plane and vertical growth mode on nonreactive substrates by physisorption. The adsorption of precursor molecules on the basal plane of grown SnS is dominant due to their limited surface migration. (c) Basal-plane and lateral growth mode on nonreactive substrates by physisorption. The surface migration of precursor molecules on the mica substrate is enhanced due to the S-rich surface conditions.

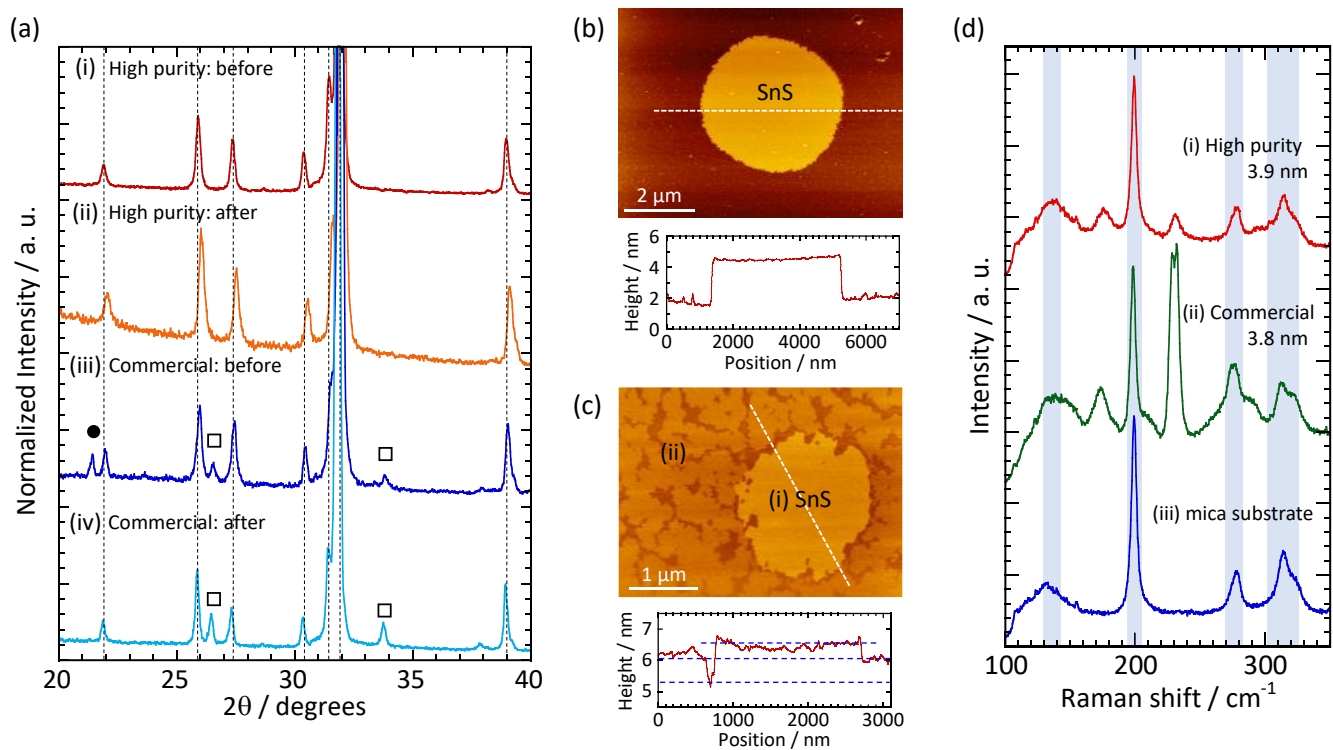

**Fig. 2** (a) XRD patterns of the high purity powder for before (i) and after (ii) the growth run and of the purchased powder for before (iii) and after (iv) the growth run. The dotted line, solid circle and open rectangles indicate SnS (JCPDS: 83-1758), $Sn_2S_3$ (JCPDS: 75-2183) and $SnO_2$ (JCPDS 41-1445), respectively. AFM images of the SnS films grown from (b) the high purity powder and (c) the commercial powder. (d) Raman spectra of the SnS films measured at 5 K. The peaks in the hatch come from the mica substrate.

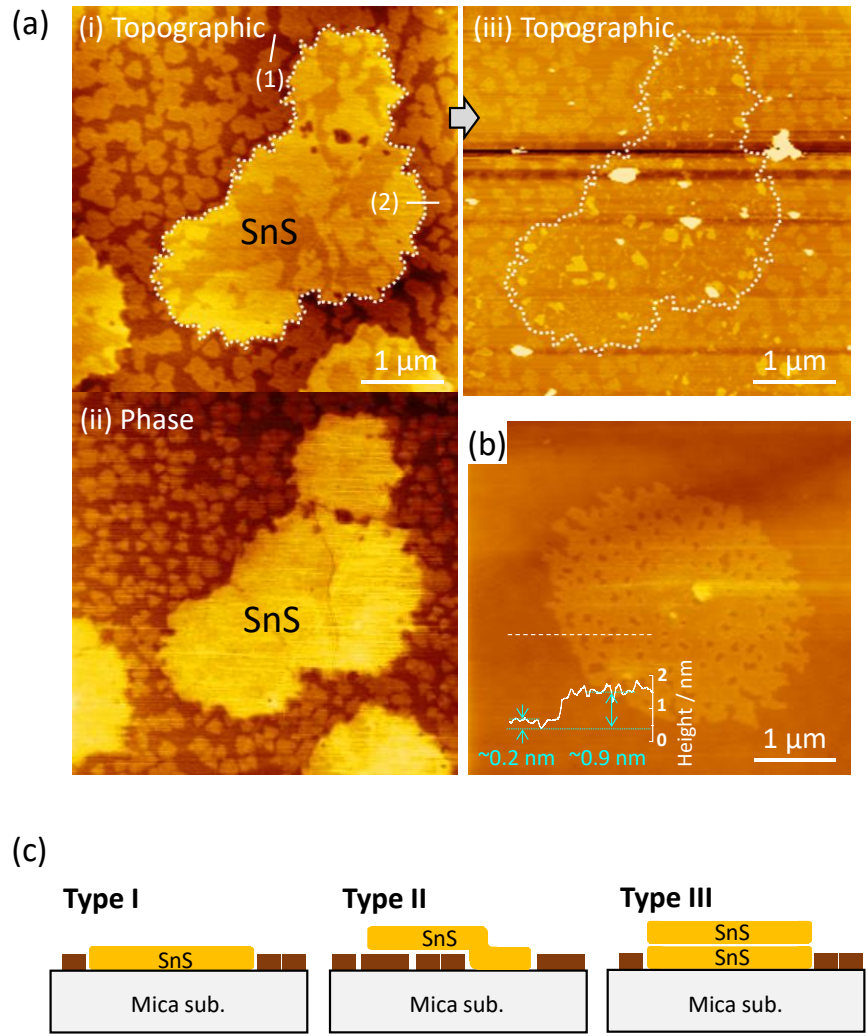

**Fig. 3** (a) AFM images of SnS film on a mica substrate. (i) Topographic image and (ii) phase image of the same position. (iii) Topographic image after the reheating. (b) AFM topographic image of monolayer SnS film. (c) Three different arrangements for the SnS film and small grains.

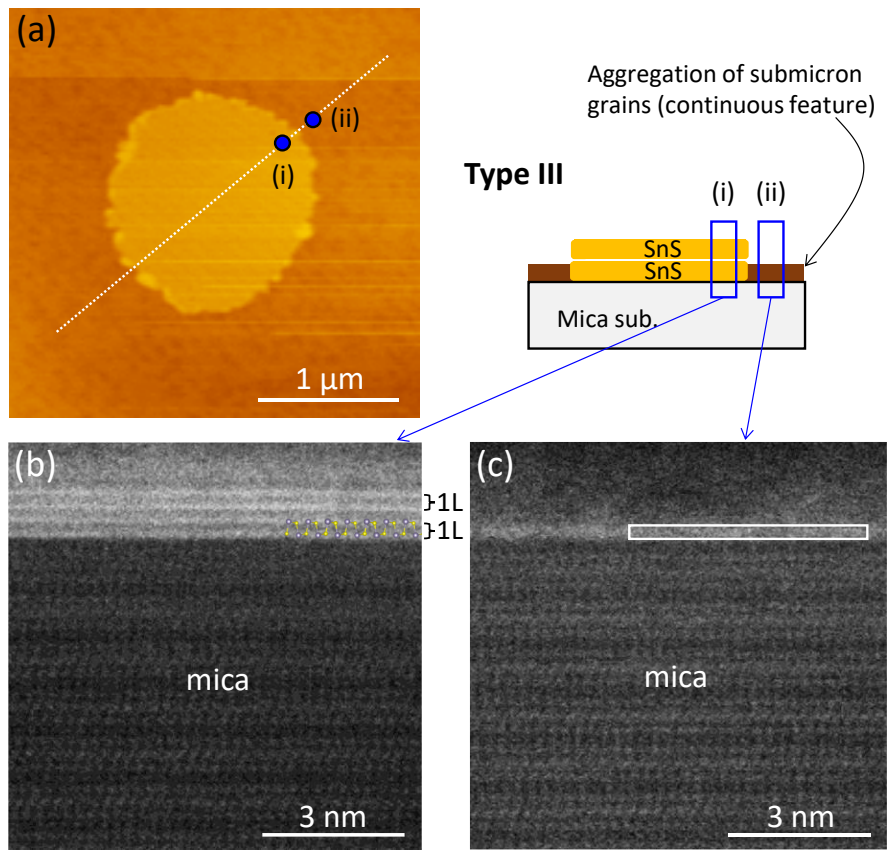

**Fig. 4** (a) AFM topographic image of SnS film. (b) Cross-sectional STEM image at position (i). Since the zone axis of SnS puckered structure could not be adjusted, the atomic configuration is not seen. However, a bilayer SnS is clearly detected. (c) Cross-sectional STEM image at position (ii). EDS was taken for the aggregation of submicron grains at the rectangular region.

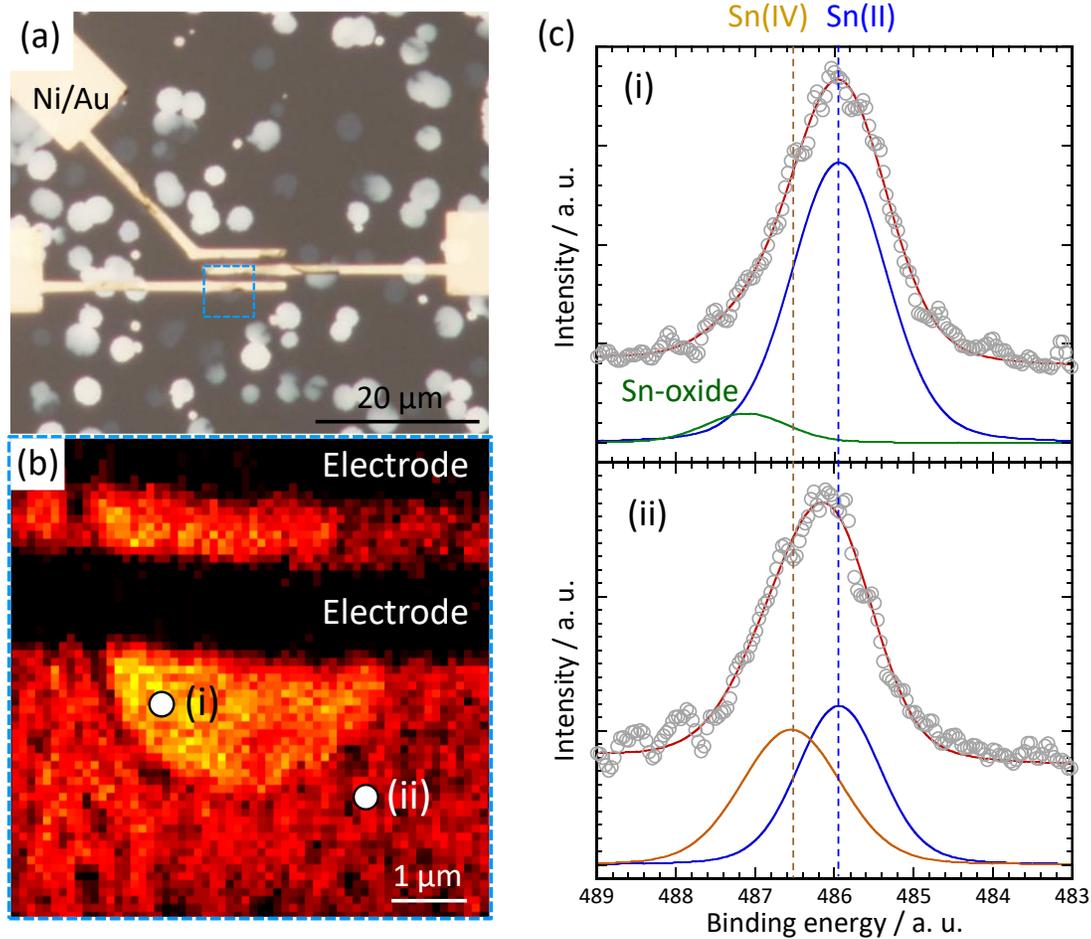

**Fig. 5** (a) Optical image of SnS film on a mica substrate connected by Ni/Au electrodes, which are grounded to the 3D-nano ESCA sample holder. (b) Photoelectron intensity mapping image for the Sn $3d_{5/2}$ peak with a spatial resolution of 100 nm. The blue dotted rectangular region in (a) was selectively measured. (c) Pinpoint Sn $3d_{5/2}$ peak core-level spectra recorded at points (i) and (ii).

**Supplementary infromation:**

## Micrometer-scale monolayer SnS growth by physical vapor deposition
H. Kawamoto[a], N. Higashitarumizu[a], N. Nagamura[b,c], M. Nakamura[b], K. Shimamura[b], N. Ohashi[b], and K. Nagashio[*a]

[a]Department of Materials Engineering, The University of Tokyo, Tokyo 113-8656, Japan
[b]National Institute for Materials Science, Ibaraki 305-0044, Japan
[c]PRESTO, Japan Science and Technology Agency, Saitama, 332-0012, Japan
[*]nagashio@material.t.u-tokyo.ac.jp

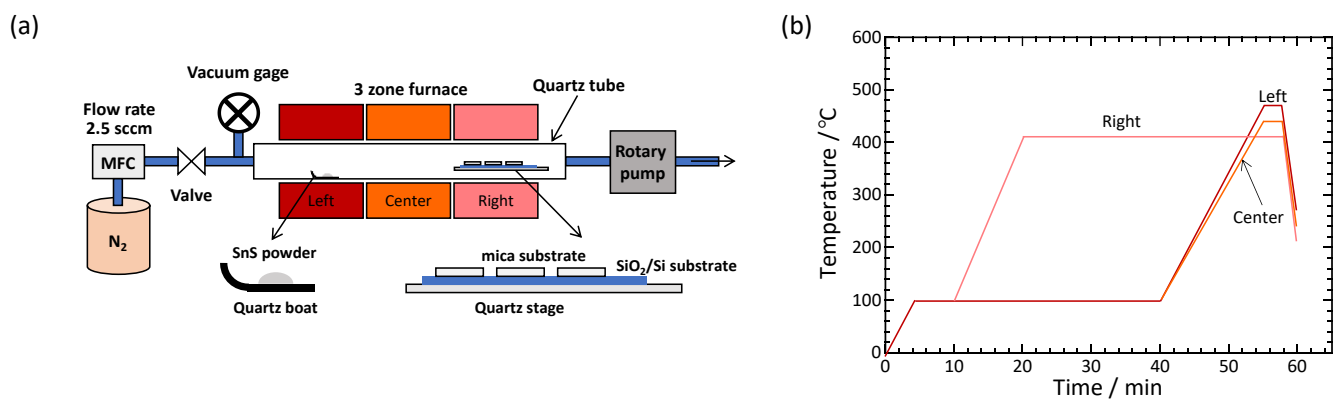

**Fig. S1.** (a) Schematic illustration of three zone furnace for PVD growth. (b) Typical temperature setup for monolayer SnS growth. The right, center, and left heater temperatures are 410°C, 440°C, and 470°C, respectively. The distance from the SnS feed powder to the mica substrate is approximately 30 cm. The pressure during the growth was kept at 10 Pa.

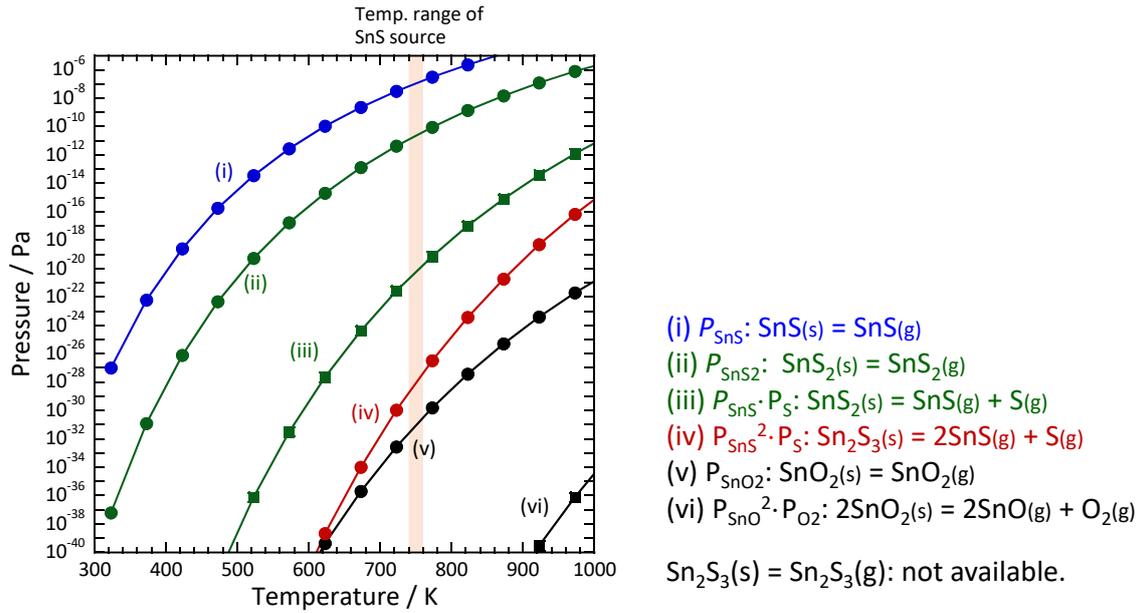

(i) $P_{SnS}$: $SnS_{(s)} = SnS_{(g)}$
(ii) $P_{SnS2}$: $SnS_{2(s)} = SnS_{2(g)}$
(iii) $P_{SnS} \cdot P_S$: $SnS_{2(s)} = SnS_{(g)} + S_{(g)}$
(iv) $P_{SnS}^2 \cdot P_S$: $Sn_2S_{3(s)} = 2SnS_{(g)} + S_{(g)}$
(v) $P_{SnO2}$: $SnO_{2(s)} = SnO_{2(g)}$
(vi) $P_{SnO}^2 \cdot P_{O2}$: $2SnO_{2(s)} = 2SnO_{(g)} + O_{2(g)}$

$Sn_2S_3(s) = Sn_2S_3(g)$: not available.

**Fig. S2.** Equilibrium vapor pressure as a function of temperature for different reactions calculated from the standard Gibbs free energy change ($\Delta G° = RT\ln K$). R is the gas constant, and $K$ is the equilibrium constant and can be approximately expressed as the partial pressure of the gas phase ($P_{gas}$). Although the product of two partial pressures is only available in reactions (iii), (iv), and (vi), the vapor pressure for each gas phase can be considered to be higher than the product pressures. Therefore, these data clearly indicate that the vapor pressures of $SnO_2(g)$ and $SnO(g)$ at the temperature for the SnS feed powder are much smaller than those of $SnS(g)$ and $S(g)$. This is consistent with the fact that $SnO_2$ remained in the purchased powder even after the growth run.

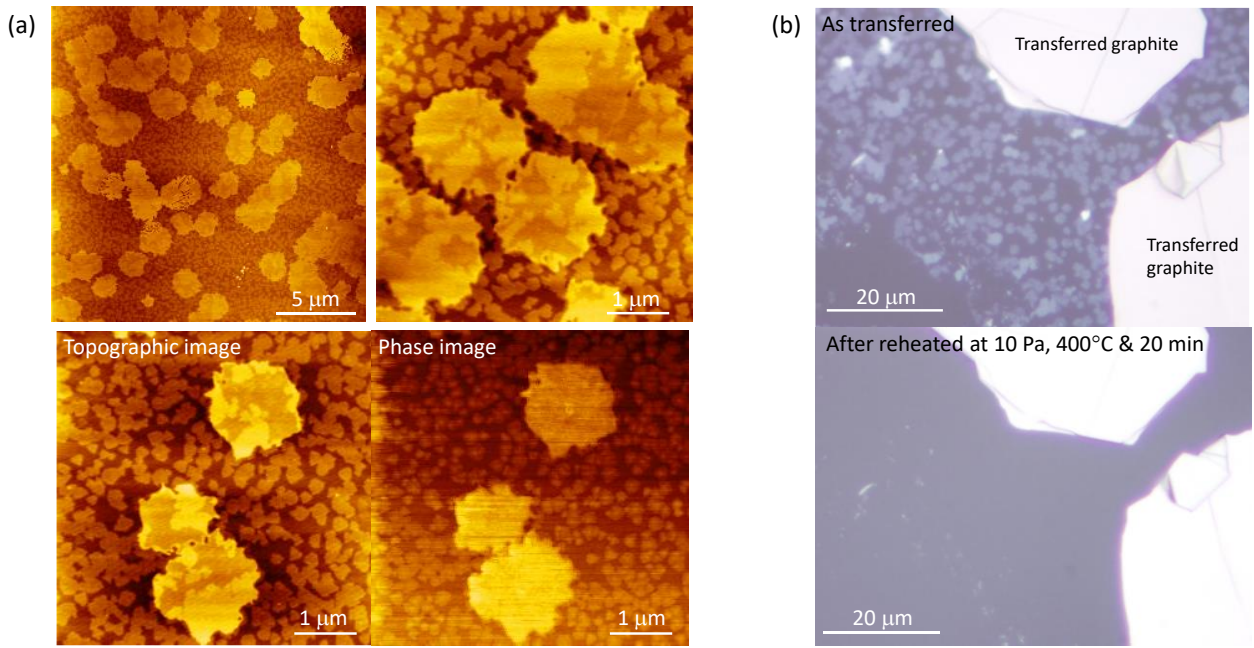

**Fig. S3.** (a) AFM images for SnS films with small grains. (b) Optical images of SnS film on a mica substrate before and after the reheating experiment. First, an AFM image of the SnS film was taken. A graphite flake was placed at the predetermined position near the desired SnS film as a position mark using a transfer system (following refs). Then, the sample was reheated in $N_2$ atmosphere at 400°C and a furnace pressure of 10 Pa for 20 min. Finally, an AFM image was taken at the same position as before.

S. Toyoda, T. Uwanno, T. Taniguchi, K. Watanabe, and K. Nagashio, *Appl. Phys. Express*, 2019, **12**, 055008.
T. Uwanno, Y. Hattori, T. Taniguchi, K. Watanabe, and K. Nagashio, *2D mater.*, 2015, **2**, 041002.

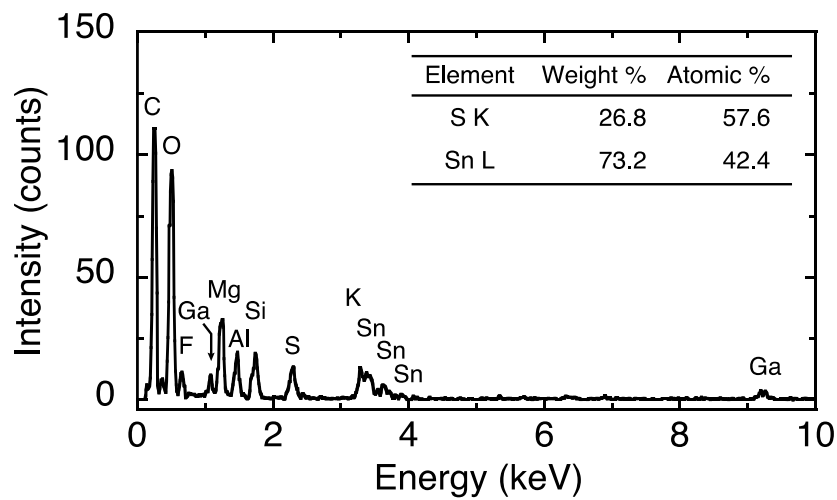

**Fig. S4.** EDS spectra for a submicron grain as highlighted in **Fig. 4c**.

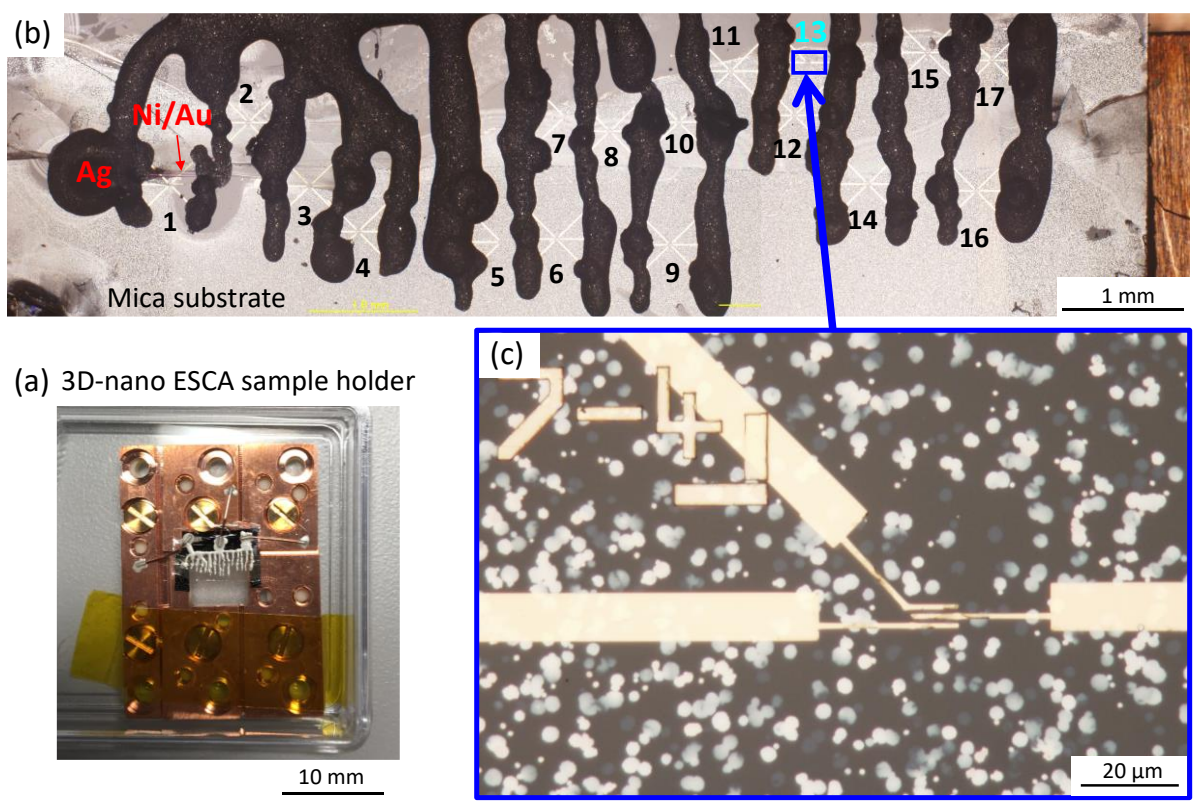

**Fig. S5.** 3D-nano ESCA installed at The University of Tokyo outstation beamline BL07LSU in SPring-8 was used for the chemical analysis. (a) Sample holder for 3D-nano ESCA, where the Ni/Au electrodes are grounded to the sample holder using Cu wire and Ag paste. (b) Low magnification optical image of the device. (c) Magnified optical image of the SnS film with Ni/Au electrodes.